\documentclass[aps,prl,twocolumn,showpacs]{revtex4}
\usepackage{graphicx}
\usepackage{wasysym}

\begin{document}
\def\be{\begin{equation}}
\def\ee{\end{equation}}

\title{Self-organized criticality model for brain plasticity}

\author{Lucilla de Arcangelis,$^1$ 
Carla Perrone-Capano$^2$ and Hans J. Herrmann$^3$}
\affiliation
{$^1$ Dept. of Information Engineering and CNISM,
Second University of Naples, 81031 Aversa (CE), Italy\\  
$^2$ Dept. of Biological Sciences, University of Naples "Federico II", 
80134, Naples, Italy and IGB "A.Buzzati Traverso", CNR, 80131 Naples, 
Italy.\\ 
$^3$ Institute for Computer Applications 1, University of Stuttgart, 
Pfaffenwaldring 27, D-70569 Stuttgart, Germany }

\begin{abstract}
Networks of living neurons exhibit an avalanche mode of activity,
experimentally found in organotypic cultures. Here we present a 
model based on self-organized criticality and taking into account brain 
plasticity, which is able to reproduce the spectrum of electroencephalograms
(EEG). The model consists in an electrical network with threshold firing and 
activity-dependent synapse strenghts. The system exhibits an avalanche 
activity power law distributed. 
The analysis of the power spectra of the electrical signal
reproduces very robustly the power law behaviour with the exponent 0.8,
experimentally measured in EEG spectra. 
The same value of the exponent is found on small-world lattices and 
for leaky neurons, indicating that universality holds for a wide class of
brain models. 
\end{abstract}

\pacs{05.65.+b, 05.45.Tp, 89.75.-k, 87.19.La}

\maketitle

\vskip 1cm

Cortical networks exhibit diverse patters of activity, including
oscillations, synchrony and waves. During neuronal activity, each neuron can
receive inputs by thousands of other neurons and, when it reaches a threshold,
redistributes this integrated activity back to the neuronal network. Recently
it has been shown that another mode of activity is neuronal avalanches, 
with a dynamics similar to self-organized criticality (SOC)
\cite{bak,jen,pacz,pacz2}, observed in organotypic cultures from
coronal slices of rat cortex \cite{beg} where neuronal avalanches are
stable for many hours \cite{beg2}.
The term SOC usually refers to a mechanism of slow
energy accumulation and fast energy redistribution driving the
system toward a critical state, where the distribution of
avalanche sizes is a power law obtained without fine tuning: no
tunable parameter is present in the model.
The simplicity of the mechanism at the basis of SOC has suggested
that many physical and biological phenomena characterized by power
laws in the size distribution, represent natural realizations of
the SOC idea. For instance, SOC has been proposed to model
earthquakes \cite{Bak,Sornette},
the evolution of biological systems \cite{Sneppen}, solar flare
occurrence \cite{solar}, fluctuations in confined plasma
\cite{plasma} snow avalanches\cite{snow} and rain fall \cite{rain}.

In order to monitor neural activities, different time series 
are usually analysed through power spectra and generally 
power-law decay is observed. 
A large number of time series analyses have been performed 
on medical data that are directly or indirectly related to brain activity. 
Prominent examples are EEG data which are used by neurologists to discern 
sleep phases, diagnose epilepsy and other seizure disorders as well as brain 
damage and disease \cite{gev,buz}. However,
the interpretation of physiological mechanisms at the basis of EEG
measurements is still controversial. 
Another  example of a physiological function which 
can be monitored by time series analysis is the human gait which is controlled 
by the brain \cite{hau}. For all these time series the power spectrum, i.e. 
the 
square of the amplitude of the Fourier transformation double logarithmically 
plotted against frequency, generally features a power law at least over one 
or two orders of magnitude with exponents between 1 and 0.7. Moreover,
experimental results show that neurotransmitter secretion rate exhibits 
fluctuations with time power law behaviour \cite{teich} and power laws are
observed in fluctuations of extended excitable systems driven by stochastic
fluctuations \cite{chia}. 
 
Here we present a model based on SOC ideas and taking into account synaptic
plasticity in a neural network. With this model we analyse the time
signal for electrical activity and compare the power spectra with EEG data.
Plasticity is one of the most astonishing properties of the
brain, occuring mostly during development and
learning \cite{alb,hen,abb}, and can be defined as the
ability to modify the structural and functional properties of synapses.
%In the mammalian central nervous
%system the refinement of neuronal connections is thought
%to occur during "critical periods" of early postnatal life,
%when circuits
%are particularly susceptible to electrical activity triggered by external
%sensory inputs.
Modifications in the strength of synapses are thought to underly memory and
learning. Among the postulated mechanisms of synaptic plasticity,
the activity dependent Hebbian plasticity constitutes the most fully
developed and influential model of how information is stored in neural circuits
\cite{heb,tsi,coo}. 
A large variety of models for brain activity has been proposed, based for
instance on the convolution of oscillators \cite{ash} or 
stochastic waiting times \cite{iva}. They are essentially abstract 
representations on a mesoscopic scale, but none of them is based on the 
behaviour of a neural network itself. In order to get real insights on the 
relation between time series and the microscopic, i.e. cellular, 
interactions inside a neural network, it is necessary to identify the 
basic ingredients of the brain activity possibly responsible for 
characteristic 
scale-free behaviour observed through the spectrum power law.
%This insight is the basis for any further understanding of 
%the diverse additional features that are observed and interpreted by 
%practitioners that analyse these time series for diagnosis. Therefore the 
%formulation of a brain model that yields the correct power spectrum 
%is of crucial importance for any further progress in the understanding of 
%the living brain.

In order to formulate a new model to study  EEG signals, we introduce 
within a SOC approach the three 
most important ingredients for neuronal activity, namely
threshold firing, neuron refractory period
and activity-dependent synaptic plasticity. 
We consider a simple square lattice of size $L\times L$ on which each site 
represents the cell body of a neuron, each bond a synapse. Therefore, on 
each site we have a potential $v_i$ and on each bond a conductance $g_{ij}$. 
Whenever at time $t$ the 
value of the potential at a site $i$ is above a certain threshold 
$v_i \geq v_{\rm max}$, 
approximately equal to $-55mV$ for the real brain, the neuron fires, i.e. 
generates an "action potential", distributing charges to its connected 
neighbours in proportion to the current flowing through each bond
$$v_j(t+1)=v_j(t)+ v_i(t) {i_{ij}(t)\over\sum_k i_{ik}(t)} \eqno (1)$$
where $v_j(t)$ is the potential at time $t$ of site $j$, nearest neighbor of 
site $i$, $i_{ij}= g_{ij} (v_i-v_j)$ and the sum is extended to all nearest 
neighbors $k$ of site $i$ that are at a potential $v_k < v_i$. After firing 
a neuron is set to a zero resting potential. 
The conductances are initially all set equal to unity whereas the neuron 
potentials are uniformly distributed random numbers between 
$v_{\rm max} - 2$ and $v_{\rm max} - 1$. 
The potential is fixed to zero at top and bottom whereas periodic
boundaries are imposed in the other direction.

The external stimulus is imposed at one input site in the centre of the 
lattice, and the electrical activity is monitored as function of time by 
measuring the total current flowing in the system. The firing rate of real 
neurons is limited by the refractory period, i.e. the brief period after the 
generation of an action potential during which a second action potential is 
difficult or impossible to elicit. The practical implication of refractory 
periods is that the action potential does not propagate back toward the 
initiation point and therefore is not allowed to reverberate between the cell 
body and the synapse. In our model, once a neuron fires, it remains quiescent 
for one time step and it is therefore unable to accept charge from firing 
neighbours. This ingredient indeed turns out to be crucial for a controlled 
functioning of our numerical model. In this way an avalanche of charges can 
propagate far from the input through the system.

\begin{figure}
\includegraphics[width=7.5cm]{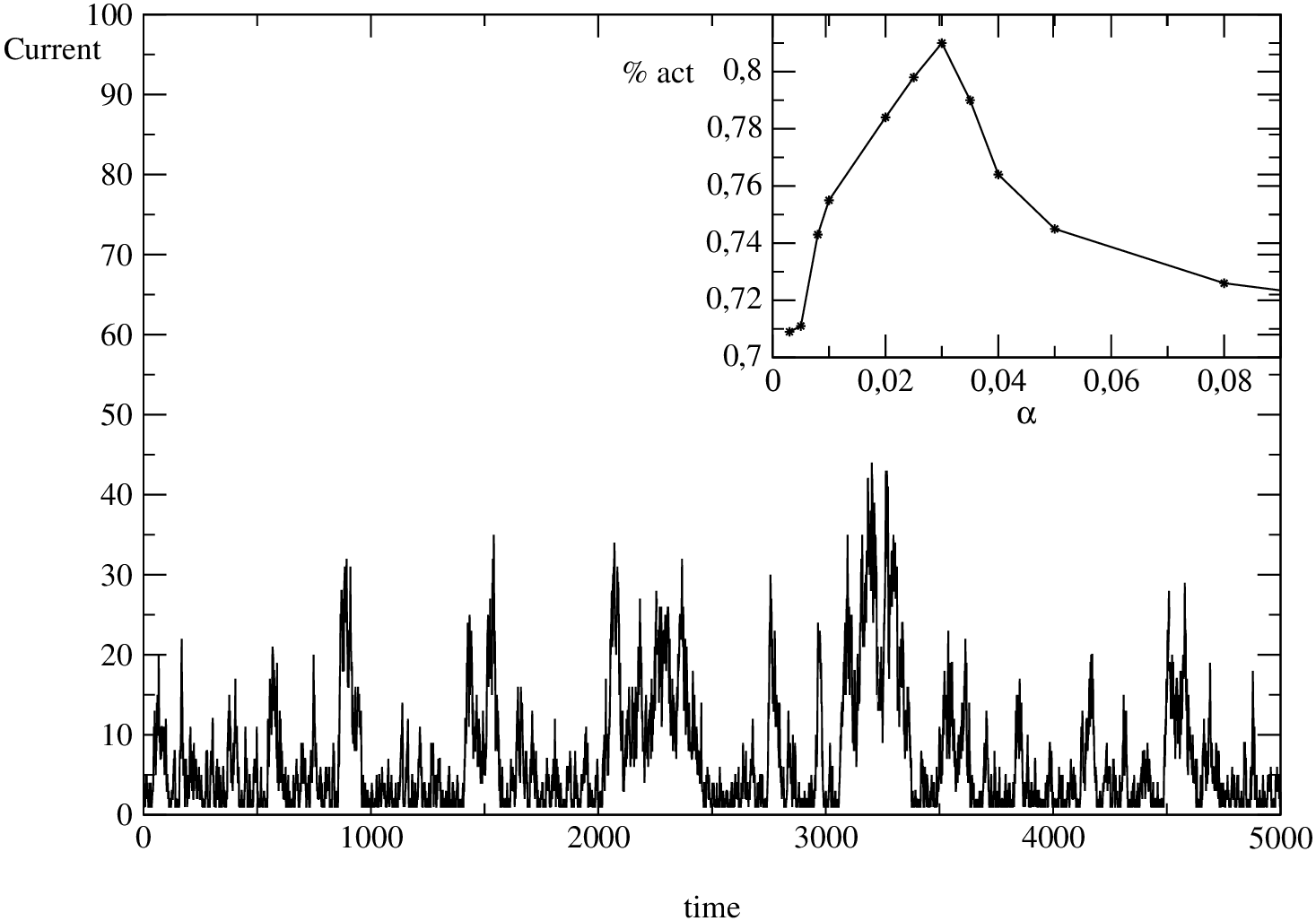}
% Here is how to import EPS art
\caption{ Total current flowing in one square lattice configuration  
($L = 1000$, $\alpha = 0.03$, $\sigma_t = 0.0001$, $v_{\rm max} = 6$)
as function of time in a sequence of several thousand stimuli. 
In the inset we show the asymptotic
value of the percentage of active bonds as function of $\alpha$ for
$L=100$. The value of the parameters is
$\sigma_t = 0.0001$ and $v_{\rm max} = 6$.
}
\label{fig1}
\end{figure}

As soon as a site is at or above threshold 
$v_{\rm max}$ at a given time $t$, it
fires according to Eq. (1), then the conductance of all the bonds, connecting
to active neurons and that have 
carried a current, is increased in the following way
$$g_{ij}(t+1) =g_{ij}(t) +\delta g_{ij} (t) \eqno (2)$$
where $\delta g_{ij}(t)=k \alpha  i_{ij}(t)$, with $\alpha$ being a 
dimensionless parameter and $k$ a unit constant bearing the dimension 
of an inverse potential. After applying Eq. (2) the time variable of our 
simulation is increased by one unit. 
Eq. (2) describes the mechanism of increase of synaptic strength,
tuned by the parameter $\alpha$. This parameter then represents the ensemble of
all possible physiological factors influencing synaptic plasticity, many of 
which are not yet fully understood.
 
Once an avalanche of firings comes to an end, the 
conductance of all the bonds with non-zero conductance is 
reduced by the average conductance increase per bond, 
$\Delta g = \sum_{ij, t} \delta g_{ij} (t)/ N_b$,
where $N_b$ is the number of bonds with non-zero conductance. 
The quantity $\Delta g$
depends on $\alpha$ and on the response of the brain to a given 
stimulus. In this way our electrical network "memorizes" the most used paths 
of discharge by increasing their conductance, whereas the less used synapses 
atrophy. Once the conductance of a bond is below an assigned small value 
$\sigma_t$, we remove it, i.e. set it equal to zero, which corresponds to 
what is known as pruning. This remodelling of synapses mimicks the fine 
tuning of wiring that occurs during "critical periods" 
in the developing brain, when neuronal activity 
can modify the synaptic circuitry, once the basic patterns of brain wiring 
are established \cite{hen}.
These mechanisms correspond to a Hebbian form of activity dependent
plasticity, where the conjunction of activity at the presynaptic and 
postsynaptic neuron modulates the efficiency of the synapse \cite{coo}. 
To insure the stable functioning of neural circuits, both strengthening and
weakening rule of Hebbian synapses are necessary to avoid instabilities due to
positive feedback \cite{des}. However, differently from the
well known Long Term Potentiation (LTP) and Long Term Depression (LTD)
mechanisms, in our model the modulation of synaptic strength does not depend on
the frequency of synapse activation \cite{alb,pau,bra}. 
%It should be also
%considered that in the living brain synapses exhibiting plasticity are
%not electrical but chemical. For instance,
%hebbian plasticity at excitatory synapses is classically mediated by
%postsynaptic calcium dependent mechanisms \cite{bi01}. In our approach the
%excitability of the postsynaptic neuron is simply modulated by the value of the
%electrical potential.

The external driving mechanism to the system is imposed by setting the 
potential of the input site to the value $v_{\rm max}$, corresponding to one 
stimulus. This external stimulus is
needed to keep functioning the system and therefore mimicks the
living brain activity. We let the discharge 
evolve until no further firing occurs, then we apply the next stimulus. 
Fig.1 shows the electrical signal as function of time: the total current 
flowing in the system is recorded in time during a sequence of successive 
avalanches. 
%As defined above the time unit corresponds to the time necessary to propagate 
%the signal from a neuron to next nearest neighbours. 
Data show that discharges of all sizes are present in the brain response, as in
self-organized criticality where the avalanche size distribution 
scales as a power law \cite{beg,nota}
%\begin{figure}
%\includegraphics[width=5.5cm,angle=270]{fig2b.eps}% Here is how to import EPS art
%\caption{ The average number of pruned bonds $N_{pb}$ as function of time 
%in a lattice of linear size $L = 100$
%for different values of $\alpha$. The value of the parameters is
%$\sigma_t = 0.0001$ and $v_{\rm max} = 6$. In the inset we show the asymptotic
%value of the percentage of active bonds as function of $\alpha$.
%}
% \label{fig2}
% \end{figure}
The strength of the 
parameter $\alpha$, controlling both the increase and decrease of synaptic
strength, determines the plasticity dynamics in the network. 
%In fact, 
%the more the system learns strengthening the used synapses, the more the 
%unused connections will weaken. 
%The number of pruned bonds in 
%the system as function of time indicates that, 
For large values of $\alpha$ the system strengthens more 
intensively the synapses carrying current but also very rapidly prunes the 
less used connections, reaching after a short transient a plateau where it 
prunes very few bonds. On the contrary, for small values of $\alpha$
the system takes more time to initiate the pruning process and 
slowly reaches a plateau. 
%The inset of Fig.1 shows the asymptotic value of
%the fraction of active bonds, calculated as the total number of bonds in the
%unpruned network minus the asymptotic number of pruned bonds, as function of
%$\alpha$. 
The number of active (non-pruned) bonds asymptotically reaches its largest 
value at the value $\alpha=0.03$ (inset of Fig.1). 
This could be interpreted as an optimal value for the 
system with respect to plastic adaptation.

\begin{figure}
\includegraphics[width=7.5cm]{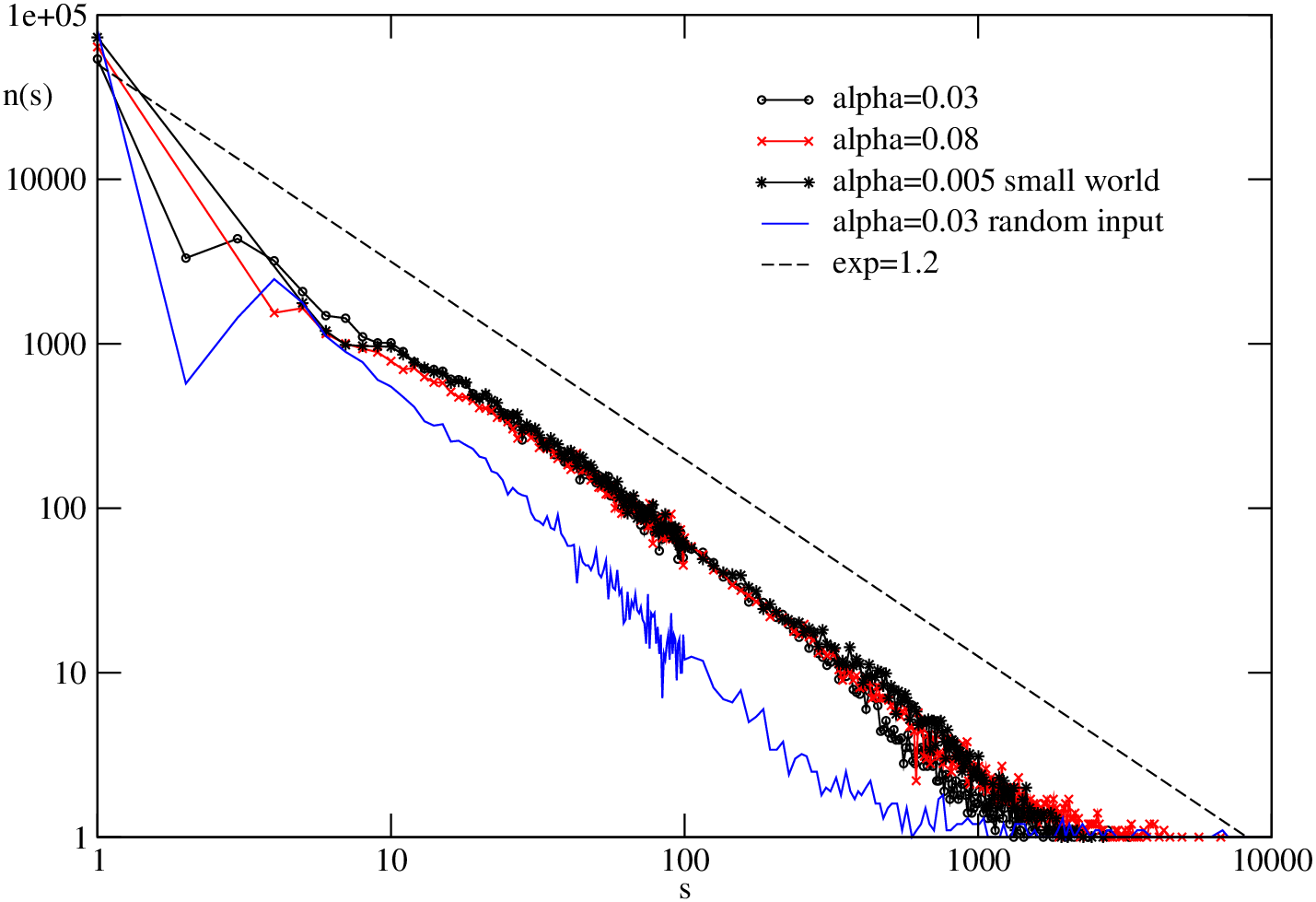}
% Here is how to import EPS art
\caption{(Color online) Log-log plot of the distribution of avalanche size
$n(s)$ ($L = 1000$,
$\alpha= 0.03$ and $0.08$, $N_p = 10$, $v_{\rm max} = 6$) for the square
lattice (lines) and the
small world lattice (*, $L = 1000$, $\alpha= 0.05$, $N_p = 1000$,
$v_{\rm max} = 8$) with $1 \%$ rewired bonds.
The data are averaged over 10000 stimuli in 10 different
configurations. The dashed line has a slope 1.2. For $\alpha=0.3$ and random
input site the slope is 1.5.
%For 5 \% of inhibitory synapses the behaviour
%becomes exponential.
}
 \label{fig2}
\end{figure}

Since each avalanche may trigger the
activity of a high number of neurons, large currents flow through the system, 
therefore after $N_p$ stimuli the network is no longer a simple square lattice 
due to pruning, but exhibits a ladder-like pattern with few lateral
connections. This complex structure  constitutes the first 
approximation to a trained brain, 
on which measurements are performed. These consist of a new 
sequence of stimuli at the input site, by setting the 
voltage at threshold, during which we measure the number of firing neurons 
as function of time. This quantity corresponds to the total current flowing in 
a discharge measured by the electromagnetic signal of the EEG. 
We have evaluated the size distribution of neural avalanches, that is the 
total number of neurons involved in the propagation of each stimulus. This
distribution exhibit power law behaviour, with an exponent
equal to $1.2 \pm 0.1$, quite stable with respect to parameters (Fig.2).
We have also simulated the brain dynamics on a square lattice with a
small fraction of bonds, from 0 to $10 \%$, rewired to long range connections
corresponding to a small world network \cite{wat,lag,she}, which more
realistically reproduces the connections in the real brain.
Fig.2 shows the size distribution scaling with an exponent $1.2 \pm 0.1$ for a
system with $1 \%$ rewired bonds and a different set of parameters $\alpha$,
$N_p$, $v_{\rm max}$. Conversely, for the input site chosen at 
random in the system,  
the scaling exponent changes and becomes $1.5 \pm 0.1$ (Fig.2).

In order to 
compare with medical data, we calculate the power spectrum of the resulting 
time series, i.e. the square of the amplitude of the Fourier transform as 
function of frequency.
The average power spectrum as function of frequency is shown in a log-log-plot 
with the parameters $\alpha= 0.03$, $N_p  = 10$, $\sigma_t = 0.0001$, 
$v_{\rm max} = 6$ and a lattice of size $L = 1000$ (Fig. 3). We 
see that it exhibits a power law behaviour 
with the exponent $0.8 \pm 0.1$ over more than three orders of magnitude.
This is precisely the same value for the exponent 
found generically on medical EEG power spectra \cite{fre,nov}. 
We also show in 
Fig. 3 the magnetoelectroencephalography (similar to EEG) obtained from 
channel 17 in the left hemisphere of a male 
subject, as measured in ref.\cite{nov}, having the exponent 0.795.
 
\begin{figure}
\includegraphics[width=7.5cm]{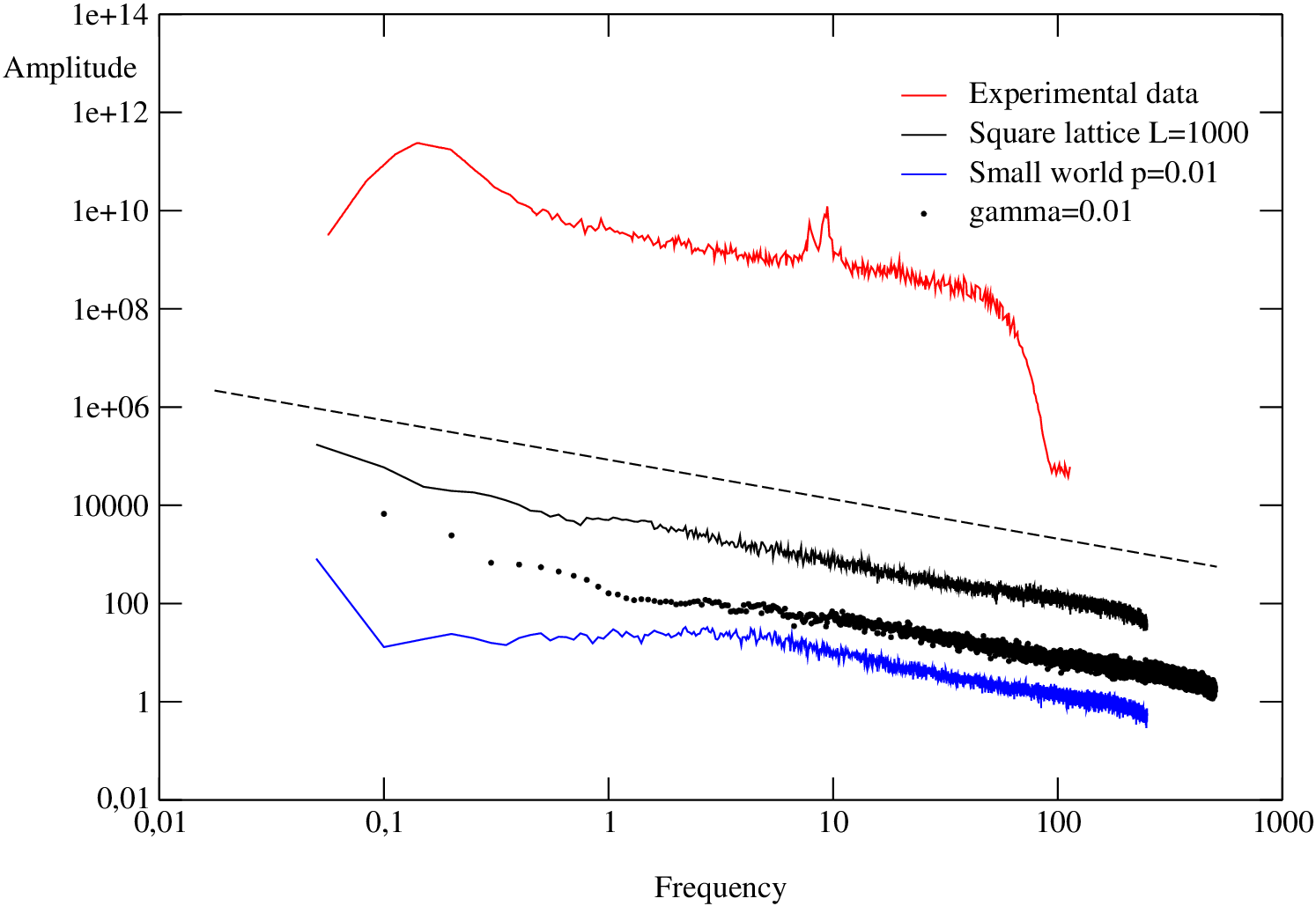}
% Here is how to import EPS art
\caption{(Color online) Power spectra for experimental data and numerical data 
($L = 1000$, 
$\alpha= 0.03$, $N_p = 10$, $v_{\rm max} = 6$) for the square lattice 
(middle line) and the 
small world lattice (bottom line, $L = 1000$, $\alpha= 0.05$, $N_p = 1000$, 
$v_{\rm max} = 8$) with $1 \%$ rewired bonds. Spectrum for the square lattice 
$\alpha =0.3$ and leaky neurons (*,$\gamma =0.01$). The experimental data 
(top line) are from ref.[19] and frequency is in $Hz$. 
The numerical data are averaged over 10000 stimuli in 10 different network 
configurations. The dashed line has a slope 0.8.
}
 \label{fig3}
\end{figure}

We have checked that the value of the exponent is 
stable against changes of the parameters $\alpha$, $v_{\rm max}$, $\sigma_t$, 
and $N_p$, and also for random initial bond conductances.
Moreover, the scaling behaviour remains unchanged if the input site is 
placed at random in the system at each stimulus.  
For $\alpha = 0$ the frequency range of
validity of the power law decreases by more than an order of magnitude.
Fig.3 also shows the power spectrum for a small world network 
with $1 \%$ rewired bonds and a different set of parameters $\alpha$, 
$N_p$, $v_{\rm max}$: the spectrum has some deviations from the power law at 
small frequencies and tends to the same universal scaling behaviour at larger 
frequencies over two orders of magnitude. The same behaviour is found for a 
larger fraction of rewired bonds. 
 
In real systems neurons have a leakage, namely the potential decays
exponentially in time with a relaxation time $\tau$, i.e. 
${dv(t)\over dt}=-\gamma v(t)$, with $\gamma=1/\tau$. Leakage has been 
considered in our model and
the same scaling behaviour recovered (Fig.3). However for $\tau \leq 10$ 
(i.e. for stronger leaking), the low frequency part of the spectrum 
appears to be frequency independent and the scaling regime is recovered at high
frequencies with an exponent in agreement with previous results.

In the mature living brain synapses can be excitatory or inhibitory, namely they
set the potential of the post-synaptic membrane to a level closer or farther,
respectively, to the firing threshold. We have introduced in our model this
ingredient: each synapse is inhibitory with probability $p_{in}$ and excitatory
with probability $1-p_{in}$. We have studied the power spectrum for a range of
value of $p_{in}$. For a density up to 10 \% of inhibitory synapses the same 
power law behaviour is recovered within error bars. For increasing
density the scaling behaviour is progressively lost and the spectrum develops a
complex multi-peak structure for $p_{in}=0.5$. Furthermore, the size 
distribution exhibits an exponential behaviour even for very small 
densities of inhibitory synapses. These results suggest that the
balance between excitatory and inhibitory synapses has a crucial role on the
overall behaviour of the network, similarly to what can occur in some severe
neurological and psychiatric disorders \cite{pow}.  

The stability of the spectrum exponent suggests that an universal 
scaling characterizes a large class of brain models and physiological signal 
spectra for brain controlled activities.
Medical studies of EEG focus on subtle details of a power spectrum (e.g. shift 
in peaks) to discern between various pathologies. These detailed structures 
however live on a background power law spectrum that shows universally an 
exponent of about 0.8, as measured for instance in refs. 
\cite{fre} and \cite{nov}. A similar 
exponent was also detected in the spectral analysis of the stride-to-stride 
fluctuations in the normal human gait which can directly be related to 
neurological activity \cite{hau}. 
Our simple model is based on SOC ideas: the threshold dynamics ensures 
time scale separation (slow external drive and fast internal relaxation).
This dynamics leads to criticality and therefore power law behaviour 
\cite{jen}. 
However the new ingredients of the
model, namely the plasticity of the synapses may be at the
origin of the new observed exponent.
This work may open new perspectives to study pathological features
of EEG spectra by including further realistic details into the neuron and
synapsis behaviour.

%\begin{figure}
%\includegraphics[width=6cm,angle=270]{figleak.eps}
% Here is how to import EPS art
%\caption{Power spectra for numerical data (square lattice $L = 1000$,
%$\alpha= 0.03$, $N_p = 10$, $v_{\rm max} = 6$) for the case of leaky neurons
%with two different values of $\gamma$.
%Data are averaged over 10000 stimuli in 10 different network
%configurations.
%}
% \label{fig4}
%\end{figure}

{\small Acknowledgements. We gratefully thank E. Novikov and collaborators for 
allowing us to use their experimental data. We also thank Salvatore Striano, 
MD, for discussions and Stefan Nielsen and Hansj\"org Seybold for help.
L.d.A. research is supported by EU Network 
MRTN-CT-2003-504712, MIUR-PRIN 2004, MIUR-FIRB 2001, CRdC-AMRA,
INFM-PCI.
}

\end{document}